\documentclass[twocolumn,floatfix,aps]{revtex4}
\usepackage{graphicx}
\usepackage{amsmath}
\usepackage{amssymb}
\usepackage{bm}

\begin{document}

\title{Magnetic breakdown and Klein tunneling in a type-II Weyl semimetal}
\author{T. E. O'Brien}
\affiliation{Instituut-Lorentz, Universiteit Leiden, P.O. Box 9506, 2300 RA Leiden, The Netherlands}
\author{M. Diez}
\affiliation{Instituut-Lorentz, Universiteit Leiden, P.O. Box 9506, 2300 RA Leiden, The Netherlands}
\author{C. W. J. Beenakker}
\affiliation{Instituut-Lorentz, Universiteit Leiden, P.O. Box 9506, 2300 RA Leiden, The Netherlands}
\date{April 2016}
\begin{abstract}
The band structure of a type-II Weyl semimetal has pairs of electron and hole pockets that coexist over a range of energies and touch at a topologically protected conical point. We identify signatures of this Weyl point in the magnetic quantum oscillations of the density of states, observable in thermodynamic properties. Tunneling between the electron and hole pockets in a magnetic field is the momentum space counterpart of Klein tunneling at a \textit{p--n} junction in real space. This magnetic breakdown happens at a characteristic field that vanishes when the Fermi level approaches the Weyl point. Topologically distinct, connected or disconnected, pairs of type-II Weyl cones can be distinguished by the qualitatively different dependence of the quantum oscillations on the direction of the magnetic field.
\end{abstract}
\maketitle

Weyl semimetals provide a condensed matter realization of massless relativistic fermions \cite{Wey29}. Their spectrum features a diabolo-shaped surface in energy-momentum space that separates helical electron-like states (moving in the direction of the momentum) from hole-like states (moving opposite to the momentum) \cite{Her37}. These ``Weyl cones'' are the three-dimensional analogue of the two-dimensional Dirac cones in graphene. The third spatial dimension provides a topological protection, by which the conical point (Weyl point) cannot be opened up unless two Weyl cones of opposite helicity are brought together in momentum space \cite{Volovik}.

Although the Weyl point cannot be locally removed, the cones can be tilted and may even tip over \cite{Huh02,Kaw12,Tre15,Sun15,Xu15,ref:Bernevig,Aut16,Koe16,Mue16}. For the relativistic Weyl cone such a distortion is forbidden by particle-hole symmetry, but that is not a fundamental symmetry in condensed matter. While in graphene the high symmetry of the honeycomb lattice keeps the cone upright, strain providing only a weak tilt \cite{Goe08}, the tilting can be strong in 3D Weyl semimetals. This leads to a natural division of Weyl cones into two topologically distinct types \cite{ref:Bernevig}. In type I the cone is only weakly tilted so that the electron-like states and hole-like states occupy separate energy ranges, above or below the Weyl point. In type II the cone has tipped over so that electron and hole states coexist in energy. Many experimental realizations of a type-II Weyl semimetal have recently been reported \cite{Hua16,Xu16,Den16,Jia16,Lia16,NXu16,Wang16}.

\begin{figure}[tb]
\centerline{\includegraphics[width=0.8\linewidth]{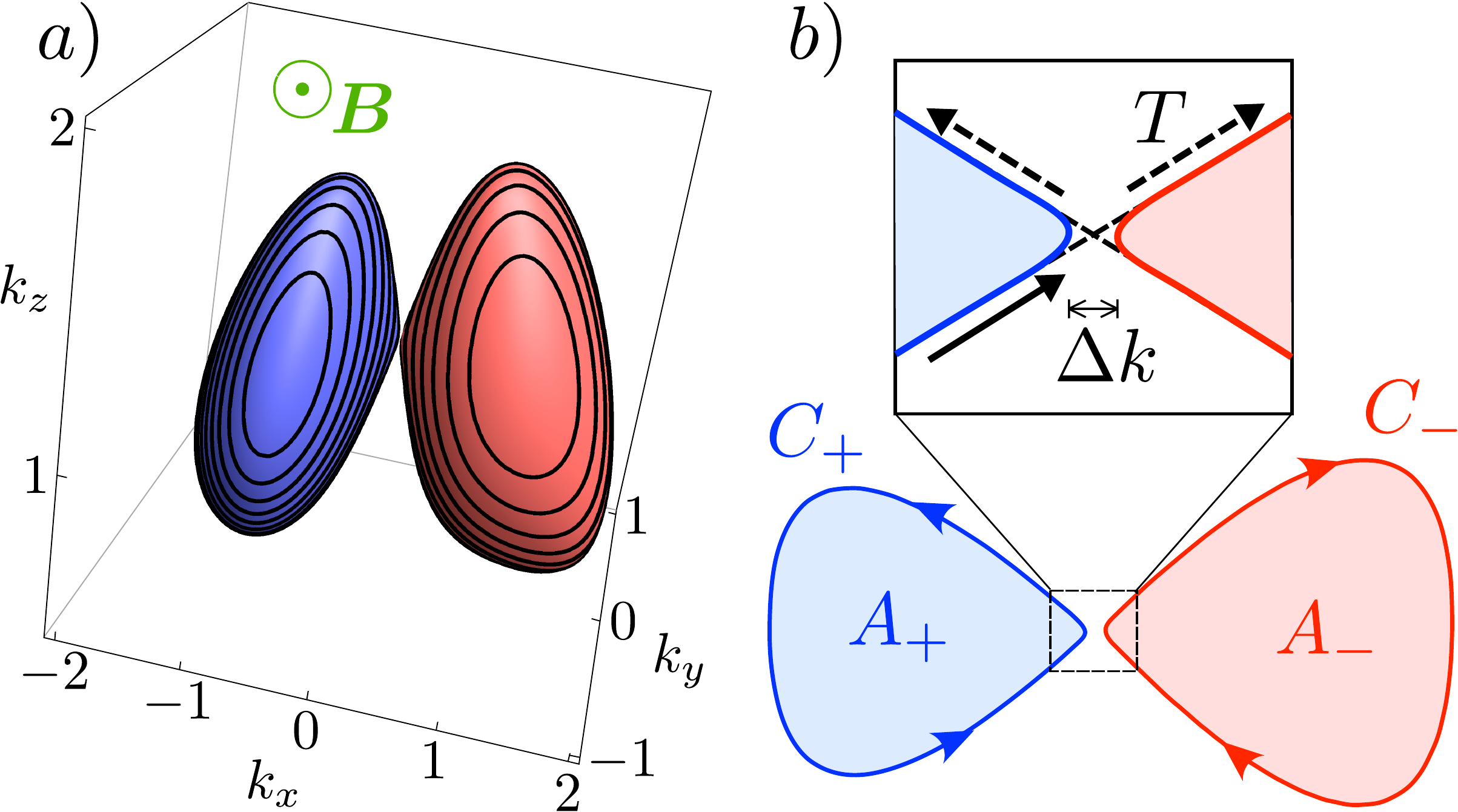}}
\caption{a) Fermi surface of a type-II Weyl semimetal, calculated from the model Hamiltonian \eqref{eqn:gen_Weyl_ham}, showing the electron and hole pockets touching at the Weyl point. Equi-energy contours in planes perpendicular to the magnetic field $B$ are indicated. The magnetic quantum oscillations have a periodicity in $1/B$ determined by the contour that encloses an extremal area.
b) Intersection of the Fermi surface with a plane perpendicular to $B$ that passes through the Weyl point. Electron and hole pockets are bounded by a contour $C_\pm$ enclosing an area $A_\pm$. The semiclassical orbit of an electron follows the contour in the direction of the arrow. Tunneling between the pockets happens with a probability $T$ that tends to unity when their minimal separation $\Delta k\rightarrow 0$. This magnetic breakdown is a manifestation of Klein tunneling in momentum space.
\label{fig:orbits}
}
\end{figure}

In a magnetic field the coexisting electron and hole pockets of a type-II Weyl semimetal are coupled by tunneling through the Weyl point (Fig.\ \ref{fig:orbits}). Here we investigate how this process, a momentum space manifestation of Klein tunneling \cite{Har09}, affects the magnetic quantum oscillations of the density of states (De Haas-Van Alphen effect), providing a unique thermodynamic signature of the topologically protected band structure (an alternative to proposed transport signatures \cite{ref:Bernevig,Cha15,Zyu16,Yu16}). Because the quantum oscillations are governed by extremal cross-sections of the Fermi surface, one might wonder whether some high symmetry is required to align the extremal cross-section with the Weyl point, so that it becomes observable. Our analysis shows that a magnetic field axis for this alignment exists generically, because of the M\"{o}bius strip topology of the projective plane. We first consider Klein tunneling through a single type-II Weyl point, and then turn to pairs of Weyl cones of opposite helicity, which can be combined in topologically distinct ways \cite{Xu15} --- with a qualitatively different dependence on the Klein tunneling probability.

To first order in momentum $\bm{k}$, the Hamiltonian of a Weyl cone has the generic form
\begin{equation}
H=\textstyle{\sum_{ij}}v_{ij}k_i\sigma_j+a_{\rm tilt} k_x\sigma_0,\label{eqn:gen_Weyl_ham}
\end{equation}
in terms of Pauli matrices $\sigma_i$, $i\in\{x,y,z\}$ (unit matrix $\sigma_0$). The eigenvalues lie on two hyperboloid sheets $E_\pm$,
\begin{equation}
E_\pm=a_{\rm tilt} k_x\pm\sqrt{\textstyle{\sum_{ijl}}v_{il}v_{jl}k_i k_j}\label{eqn:gen_Weyl_e},
\end{equation}
that touch at the Weyl point $\bm{k}=0$.

For sufficiently small $a_{\rm tilt}$ the Fermi surface contains either electron-like states in $E_+$ or hole-like states in $E_-$, depending on the sign of the Fermi energy. With increasing $a_{\rm tilt}$ the Weyl cone is tilted in the (arbitrarily chosen) $x$-direction, and when it tips over coexisting electron and hole states appear on the Fermi surface. This is the type-I to type-II Weyl semimetal transition \cite{ref:Bernevig}.

The hyperboloid dispersion \eqref{eqn:gen_Weyl_e} only holds near the Weyl point. In the physical realizations of a type-II Weyl semimetal the Fermi surface closes away from the Weyl point, forming compact electron and hole pockets. A cross-section is defined by fixing an axis (unit vector $\hat{\bm{n}}$) and choosing a coordinate $q$ along that axis. The intersection of the Fermi surface with the plane $\hat{\bm{n}}\cdot\bm{k}=q$ is an oriented contour $C_\pm(q)$ enclosing the signed area $A_\pm(q)$ (positive for $C_+$ and negative for $C_-$). The contours are the classical momentum-space orbits for a magnetic field $B$ in the $\hat{\bm{n}}$-direction, the change in orientation between $C_+$ and $C_-$ resulting from the sign change of the effective mass in the electron and hole pockets.

Semiclassical quantization of the orbits produces Landau tubes \cite{ref:Ashcroft_Mermin}, with quantized cross-sectional area
\begin{equation}
A_\pm(q)=2\pi (n+\nu)eB/\hbar,\;\;n=\pm 1,\pm 2.\label{Landautube}
\end{equation}
The Maslov index $\nu=1/2$ for massive electrons, while $\nu=0$ for massless Weyl fermions \cite{Wan16}. The Landau tubes give rise to oscillations in the density of states periodic in $1/B$ \cite{note1}, 
\begin{equation}
\delta\rho/\rho_0=\,{\rm Re}\,\left\{[-iA''_\pm(q_c)]^{-1/2}e^{ 2\pi i (F_\pm/B-\nu)}\right\},\label{deltarho}
\end{equation}
with frequency given by the Onsager relation \cite{ref:Onsager,ref:Shoenberg}
\begin{equation}
F_\pm=(\hbar/2\pi e)|{\cal A}_\pm|.
\label{eq:Onsager_formula}
\end{equation}
The extremal area ${\cal A}_\pm=A_\pm(q_c)$ is the area at which the first derivative $dA_\pm(q)/dq=0$. The contour enclosing the extremal area is denoted by ${\cal C}_\pm$. 

The two sheets $E_\pm$ of a type-II Weyl cone are coupled by quantum tunneling. This magnetic-field-induced tunneling between electron and hole pockets is the momentum space counterpart of Klein tunneling at a \textit{p--n} junction in graphene \cite{Bee08}, and can be analyzed along the same lines \cite{Nan11}.

The effect of a magnetic field $B$ in, say, the $y$-direction, with vector potential $\bm{A}=(Bz,0,0)$, is accounted for by the substitution $k_x\mapsto k_x+eBz$ (setting $\hbar=1$). In momentum representation, the Schr\"{o}dinger equation $H\psi=E\psi$ reads
\begin{subequations}
\label{M0M1def}
\begin{align}
&i U_0\frac{\partial\psi}{\partial k_z}=U(k_z)\psi,\;\;
U_0=eB\bigl(\textstyle{\sum_{j}}v_{xj}\sigma_j+a_{\rm tilt}\sigma_0\bigr),\\
&U(k_z)=E\sigma_0-\textstyle{\sum_{ij}}v_{ij}k_i\sigma_j-a_{\rm tilt}k_x\sigma_0.
\end{align}
\end{subequations}
For $a_{\rm tilt}>(\sum_j v_{xj}^2)^{1/2}$ the matrix $U_0$ is positive definite, so that it can be factorized as $U_0=VV^\dagger$ with invertible $V$ and we may write
\begin{equation}
i \partial\psi/\partial k_z=V^{-1}U(k_z)(V^\dagger)^{-1}\psi\equiv {\cal H}(k_z)\psi,
\label{calHdef}
\end{equation}
with ${\cal H}(k_z)={\cal H}_0+{\cal H}_1 k_z$. If we interpret $k_z\equiv t$ as ``time'', this looks like a Schr\"{o}dinger equation for a spin-$1/2$ particle with ``time''-dependent Hamiltonian ${\cal H}(t)$. Because the $t$-dependence of ${\cal H}(t)$ is linear, we can use the Landau-Zener formula for the tunneling probability between the electron and hole pockets \cite{Lan77}.

Quite generally, for a two-level system with time-dependent Hamiltonian
\begin{equation}
{\cal H}(t)=\begin{pmatrix}
\alpha t+c&\gamma\\
\gamma^\ast&\beta t+c'
\end{pmatrix},\label{Htgeneral}
\end{equation}
the Landau-Zener tunnel probability is
\begin{equation}
T=\exp\bigl(-{2\pi|\gamma|^2}|\alpha-\beta|^{-1}\bigr).\label{TLZ_general}
\end{equation}
The matrix \eqref{calHdef} is of the form \eqref{Htgeneral} in the basis where ${\cal H}_1$ is diagonal, so in that basis we can read off the coefficients $\alpha,\beta,\gamma$ needed to determine $T$.

\begin{figure}[tb]
\centerline{\includegraphics[width=0.8\linewidth]{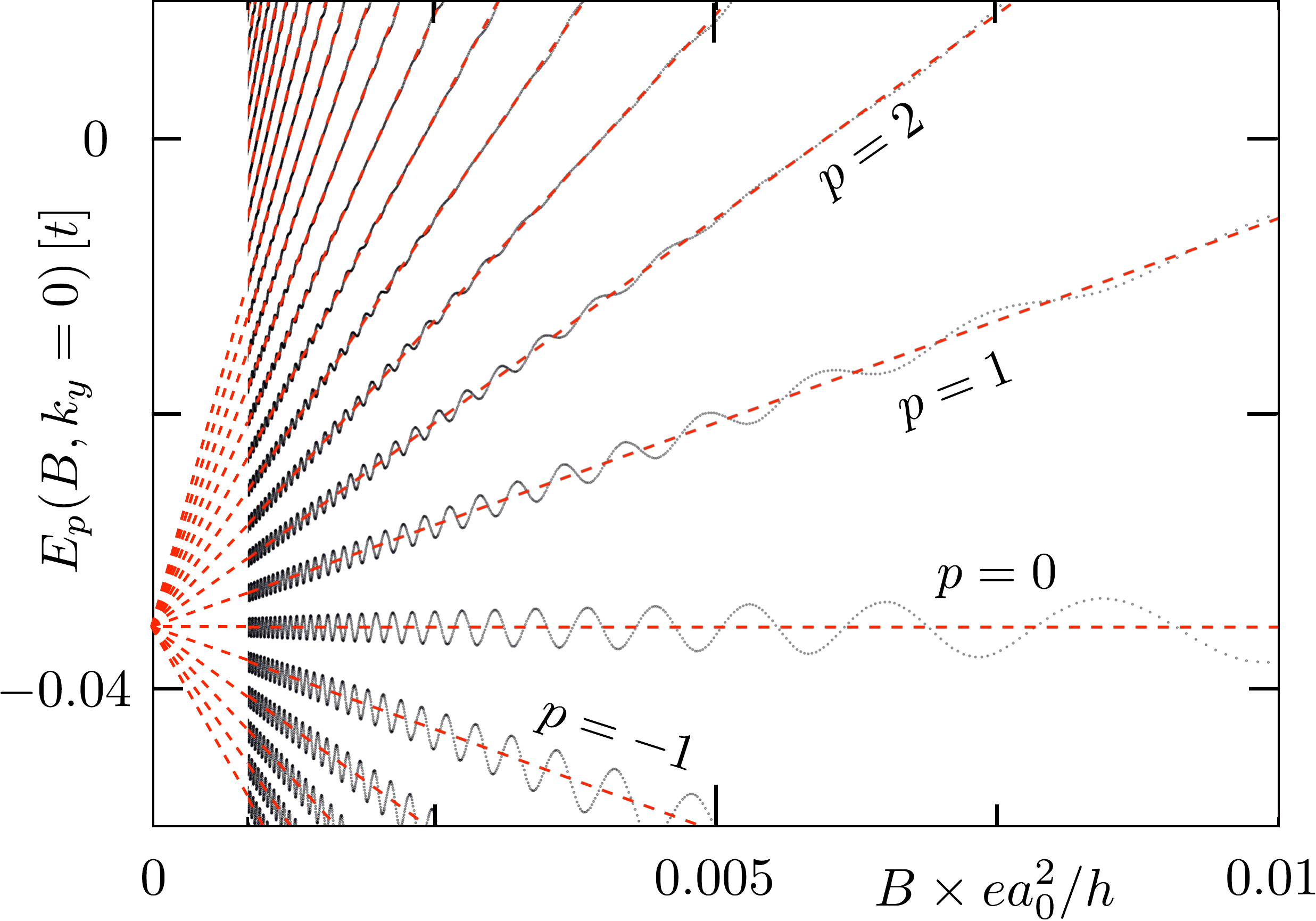}}
\caption{Energy spectrum at $k_y=0$ of the type-II Weyl semimetal with Hamiltonian \eqref{Hmodel} (parameters $t=1$, $t'=2$, $\mu=3$, $b=1.2$, $a_{\rm tilt}=1.7$, $\xi=0.08$). The black dotted curves are the exact numerical results, the red dashed lines form the semiclassical Landau fan \eqref{Landaufan} for tunnel-coupled electron and hole pockets. The individual pockets are responsible for the high-frequency oscillations superimposed on the fan.
\label{fig:Landau_levels}
}
\end{figure}

For a specific example we consider the Hamiltonian \eqref{eqn:gen_Weyl_ham} with $v_{ij}=v_i\delta_{ij}$, which for $a_{\rm tilt}>v_x$ represents a type-II Weyl cone. We find
\begin{align}
T&=\exp\left(-\frac{\pi\hbar}{eB}\right.\frac{v_x^2 E^2+v_y^2 k_y^2(a_{\rm tilt}^2-v_x^2)}{v_z(a_{\rm tilt}^2-v_x^2)^{3/2}} \left.\vphantom{\frac{\pi\hbar}{eB}}\right)\nonumber\\
&=\exp\left(-\frac{\pi\hbar}{4eBv_z}(\Delta k)^2(a_{\rm tilt}^2-v_x^2)^{1/2}\right),\label{Tresult}
\end{align}
with $\Delta k$ the minimal separation of the contours $C_+$ and $C_-$. This has the general form of the interband tunnel probability in the theory of magnetic breakdown \cite{ref:Shoenberg,ref:Blount,ref:Winkler}, with a breakdown field $B_c\propto (\Delta k)^2$. The characteristic feature of Klein tunneling is that the tunnel probability $T\rightarrow 1$ and $B_c\rightarrow 0$ at the conical point of the band structure --- here a 3D Weyl point and a 2D Dirac point in Ref.\ \onlinecite{Har09}.

To illustrate the effect of Klein tunneling between electron and hole pockets on the magnetic quantum oscillations in the density of states, we consider the model Hamiltonian \cite{Vazifeh2013}
\begin{align}
H={}&\tau_z(t'\sigma_x\sin k_x+t'\sigma_y\sin k_y )+t \tau_z\sigma_0\sin k_z\nonumber\\
&+\tau_x\sigma_0(\mu-t\cos k_x-t\cos k_y-t\cos k_z)\nonumber\\
&+b \tau_0 \sigma_z+ \bigl[a_{\rm tilt}\sin k_x +\xi(1-\cos k_x)\bigr]\tau_0\sigma_0.\label{Hmodel}
\end{align}
This is a tight-binding Hamiltonian on a cubic lattice (lattice constant $a_0=1$), with a spin and orbital degree of freedom on each lattice site (Pauli matrices $\sigma_i$ and $\tau_i$, respectively). The time-reversal symmetry breaking term $b$ splits the Dirac cone into two Weyl cones separated along the $z$-axis. To produce a type-II Weyl semimetal we have added a tilting term $a_{\rm tilt}$ and a term $\xi$ that breaks the symmetry between the electron and hole pockets.

As derived in App.\ A \cite{supplemental}, near a Weyl point the effective low-energy Hamiltonian has the form \eqref{eqn:gen_Weyl_ham} with diagonal velocity tensor $v_{ij}=v_i\delta_{ij}$ given by
\begin{subequations}
\label{vxyzresult}
\begin{align}
  v_x &= v_y = \frac{(2t-\mu)^2-t^2+b^2}{2b(2t-\mu)}t',\\
  v_z &= \frac{1}{2b} \sqrt{\left[(t-\mu)^2-b^2\right]\left[b^2-(3t-\mu)^2\right]}.
\end{align}
\end{subequations}

The Hamiltonian \eqref{Hmodel} retains a mirror symmetry in the $x$--$z$ plane (to be removed later on), which implies that for a magnetic field in the $y$-direction the areas $A_\pm(k_y)$ are extremal for $k_y=0$. By means of exact diagonalization \cite{kwant} we have calculated the partial density of states $\rho(E,B,k_y)=\sum_p \delta[E-E_p(B,k_y)]$ for $k_y=0$, assuming that this gives the dominant contribution to the magnetic quantum oscillations. We choose the gauge $\bm{A}=(0,0,-Bx)$, with a rational flux $Ba_0^2=1/N \times h/e$ through a unit cell. The lattice has dimensions $N\times NM$ in the $x$--$z$ plane ($M\gg N\gg 1$), with periodic boundary conditions in both directions.

\begin{figure}[tb]
\centerline{\includegraphics[width=0.9\linewidth]{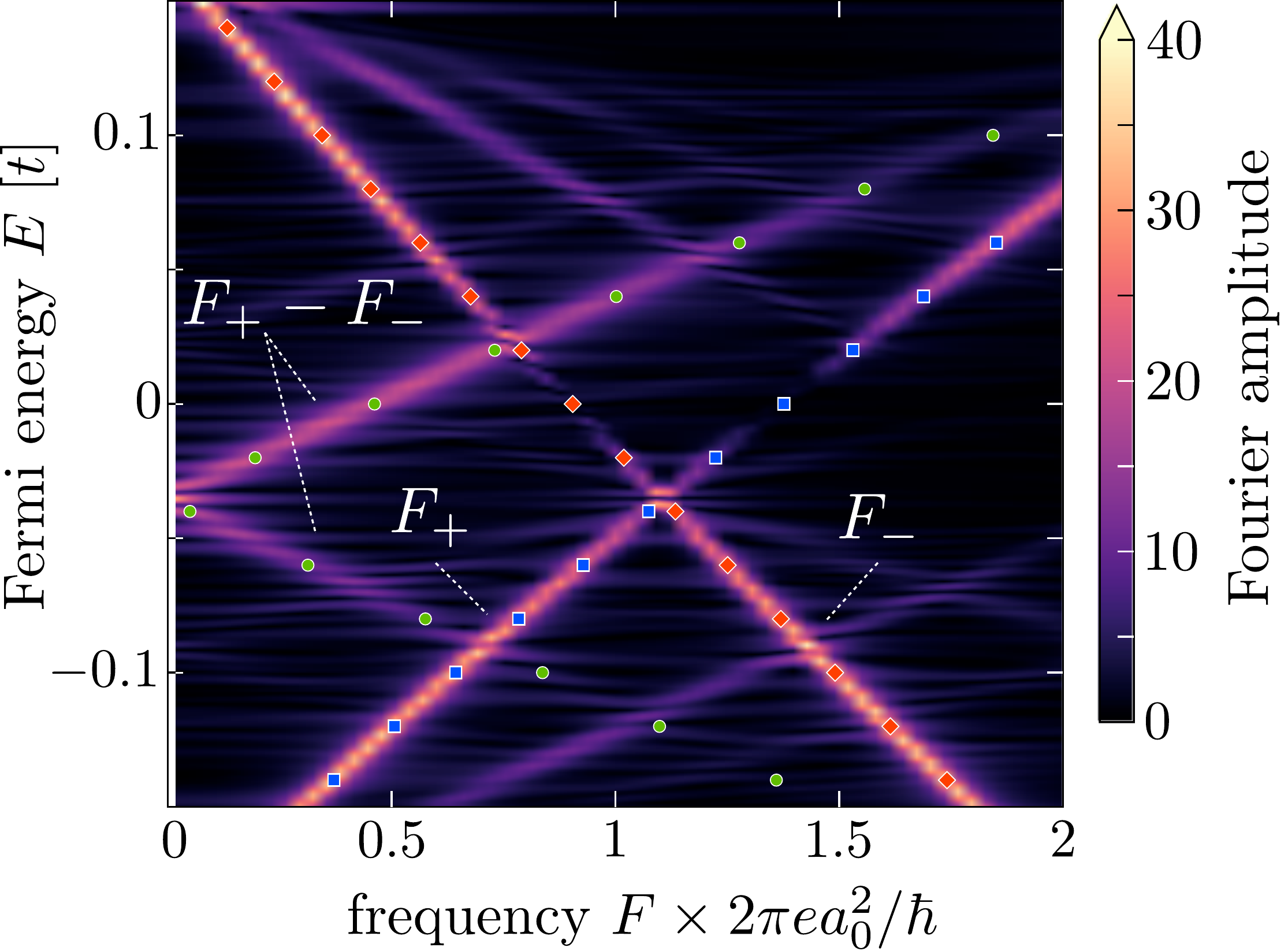}}
\caption{Fourier amplitudes of the magnetic quantum oscillations. The numerical data for the partial density of states $\rho(E,k_y=0)$ (smoothed with a Gaussian of width $\Gamma=t/500$) is Fourier transformed over the field range $B\lesssim 0.005\,h/ea_0^2$ ($200<N<1500$). The fundamental frequencies from the electron and hole pockets are indicated by $F_+$ and $F_-$, respectively (the first harmonics are also faintly visible). Klein tunneling between the pockets when the Fermi energy approaches the Weyl point ($E=0$) suppresses these high-frequency oscillations, introducing a new component at the difference frequency $|F_+-F_-|$. The colored data points for $F_\pm$ are the semiclassical prediction \eqref{eq:Onsager_formula} from the extremal areas.
\label{fig:Fourier_amplitudes}
}
\end{figure}

Fig.\ \ref{fig:Landau_levels} shows the energy spectrum as a function of magnetic field and Fig.\ \ref{fig:Fourier_amplitudes} shows the periodicity of the magnetic oscillations, extracted from a Fourier transform of the density of states. When the Fermi level is far from the Weyl point $E=0$, the electron and hole pockets contribute separately with frequencies $F_\pm$ set by the extremal areas ${\cal A}_\pm$. The slopes $dF_\pm/dE$ have opposite sign in the two pockets, signifiying the opposite sign of the cyclotron effective mass
\begin{equation}
m_{\pm}=\frac{\hbar^2}{2\pi}\frac{d}{dE}|{\cal A}_\pm|.\label{mpm}
\end{equation}

Near the Weyl point a low-frequency component appears at the difference $|F_+ -F_-|$, and the individual high-frequency components $F_\pm$ are suppressed. In a semiclassical description, the orbit responsible for the difference frequency is the ``figure of eight'' orbit formed by joining $C_+$ to $C_-$ at the Weyl point (see Fig.\ \ref{fig:orbits}b). The corresponding effective mass
\begin{equation}
m_{\Sigma}=\frac{\hbar^2}{2\pi}\frac{d}{dE}|{\cal A}_+ + {\cal A}_-| \label{mKlein}
\end{equation}
governs the Landau fan in Fig.\ \ref{fig:Landau_levels},
\begin{equation}
E_p(B)=E_p(0)+p\times\hbar eB/m_{\Sigma}. \label{Landaufan}
\end{equation}
Notice the absence of a $1/2$ offset from the integer $p$, canceled by a Berry phase.

\begin{figure}[tb]
\centerline{\includegraphics[width=0.8\linewidth]{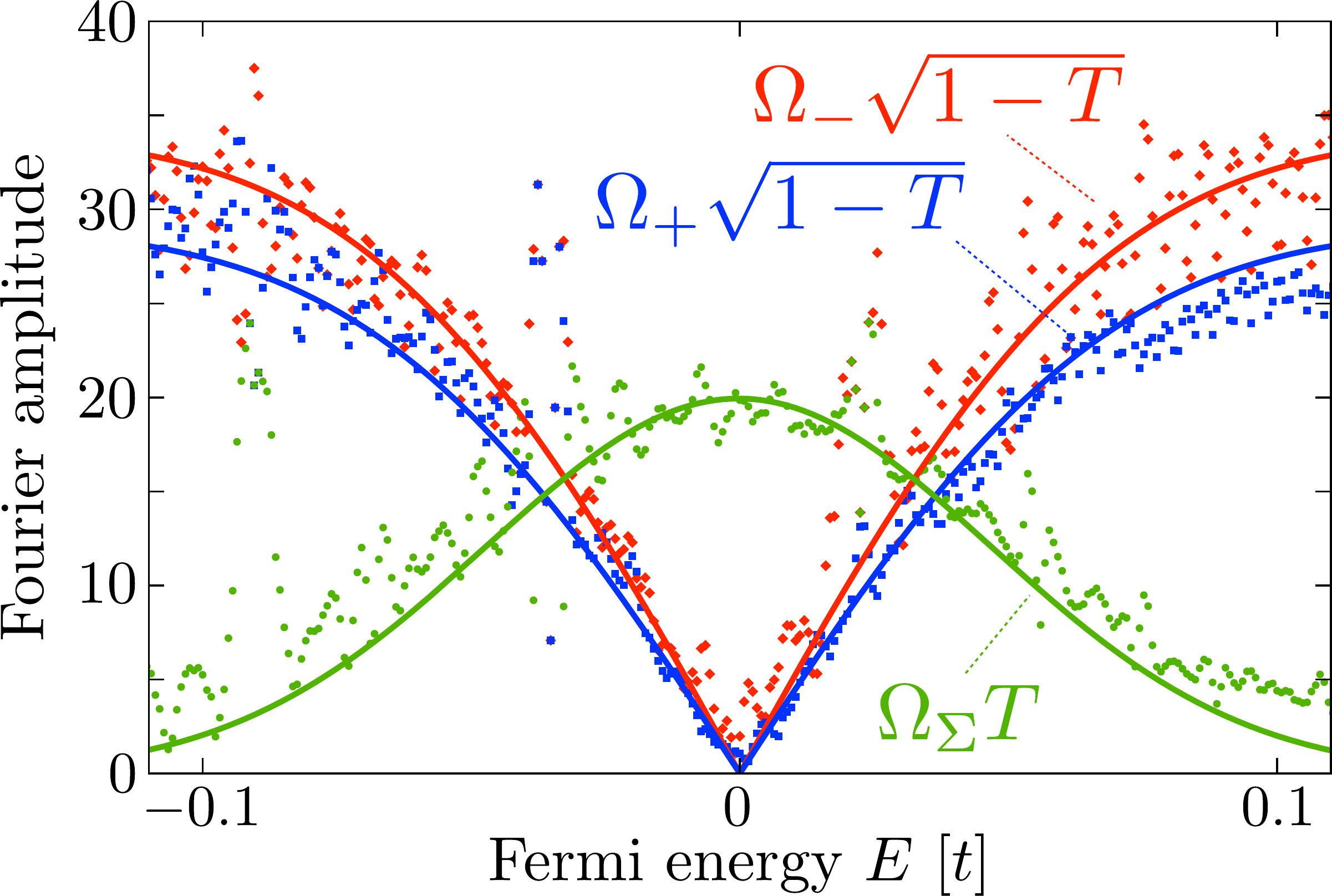}}
\caption{Energy dependence of the Fourier amplitudes from Fig.\ \ref{fig:Fourier_amplitudes}. The curves are fits to $\Omega_{\pm}\sqrt{1-T}$ and $\Omega_\Sigma T$, with the transmission probability $T(E)$ calculated from Eq.\ \eqref{Tresult} and energy-independent fit parameters $\Omega_\pm,\Omega_\Sigma$. When two frequency lines in Fig.\ \ref{fig:Fourier_amplitudes} cross we cannot reliably determine the individual amplitudes --- which explains some of the large scatter in the data points.
\label{fig:new_fits}
}
\end{figure}

The tunnel probability \eqref{Tresult} evaluates for our model parameters to $T(E)=\exp[-0.52\,N(E/t)^2]$. The contribution of an orbit to the Fourier amplitude contains a factor $t=\sqrt{T}$ for each transmission through the Weyl point and a factor $r=\sqrt{1-T}$ for each reflection. In Fig.\ \ref{fig:new_fits} we plot the peak heights of Fig.\ \ref{fig:Fourier_amplitudes} as a function of energy. The solid lines are fits to $\Omega_\pm\sqrt{1-T(E)}$ and $\Omega_\Sigma T(E)$, with energy-independent fit parameters $\Omega_\pm, \Omega_\Sigma$. We take for the inverse field strength $N = 850$, half-way the interval used in the Fourier transform. A good match to the predicted Gaussian $T(E)$ is obtained.

\begin{figure}[tb]
\centerline{\includegraphics[width=0.7\linewidth]{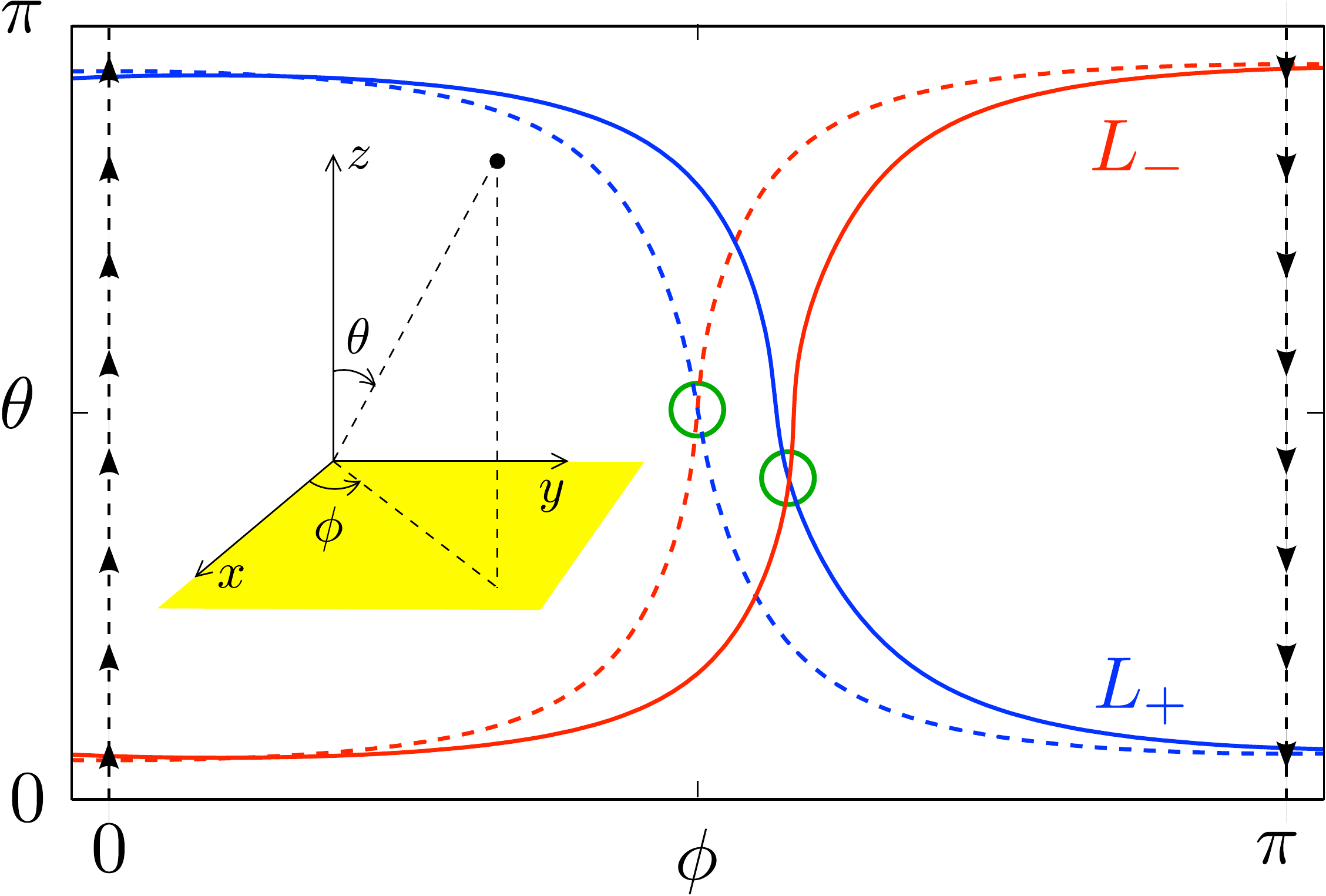}}
\caption{Magnetic field axis $L_\pm=(\theta_\pm,\phi_\pm)$ for which the extremal contour ${\cal C}_\pm$ at $E=0$ touches the Weyl point. The dashed curves correspond to the Hamiltonian \eqref{Hmodel} with the parameters of Fig.\ \ref{fig:orbits}. For the solid curves we have broken the mirror symmetry by adding the term $V_0\tau_0\sigma_0\sin k_y$ with $V_0=0.5$. The intersection of $L_+$ and $L_-$ (encircled) is the special axis at which Klein tunneling between electron and hole pockets produces magnetic quantum oscillations with the difference frequency $|F_+ -F_-|$, suppressing both the electron and hole frequencies $F_\pm$. The intersection is protected by the topology of the M\"{o}bius strip (indicated by arrows,  which show how the edges at $\phi=0,\pi$ should be glued with a twist).
\label{fig:magic_axis_fig}
}
\end{figure}

The above analysis was simplified by the mirror symmetry in the $x$--$z$ plane, because we could immediately identify the special magnetic field axis for which the extremal contours ${\cal C}_\pm$ in the electron and hole pockets both touch the Weyl point when $E\rightarrow 0$, allowing for Klein tunneling. One might wonder how restrictive this alignment is --- is it possible to find such a special axis in the absence of any symmetry? The answer is yes, as we demonstrate with the help of Fig.\ \ref{fig:magic_axis_fig}. At $E=0$ we plot the polar and azimuthal angles $\theta_\pm,\phi_\pm$ of the magnetic field axis for which the extremal contour ${\cal C}_\pm$ touches the Weyl point. Because $(\theta,\phi)$ and $(\pi-\theta,\pi+\phi)$ represent the same axis, we may restrict $\phi$ to the range $[0,\pi]$ --- half the usual range for spherical coordinates --- identifying the end points $(\theta,0)$ and $(\pi-\theta,\pi)$. The $(\theta,\phi)$ plane with these ``twisted'' periodic boundary conditions is the so-called projective plane $\mathbb{P}_2$, and has the topology of a M\"{o}bius strip.

\begin{figure}[bb]
\centerline{\includegraphics[width=0.9\columnwidth]{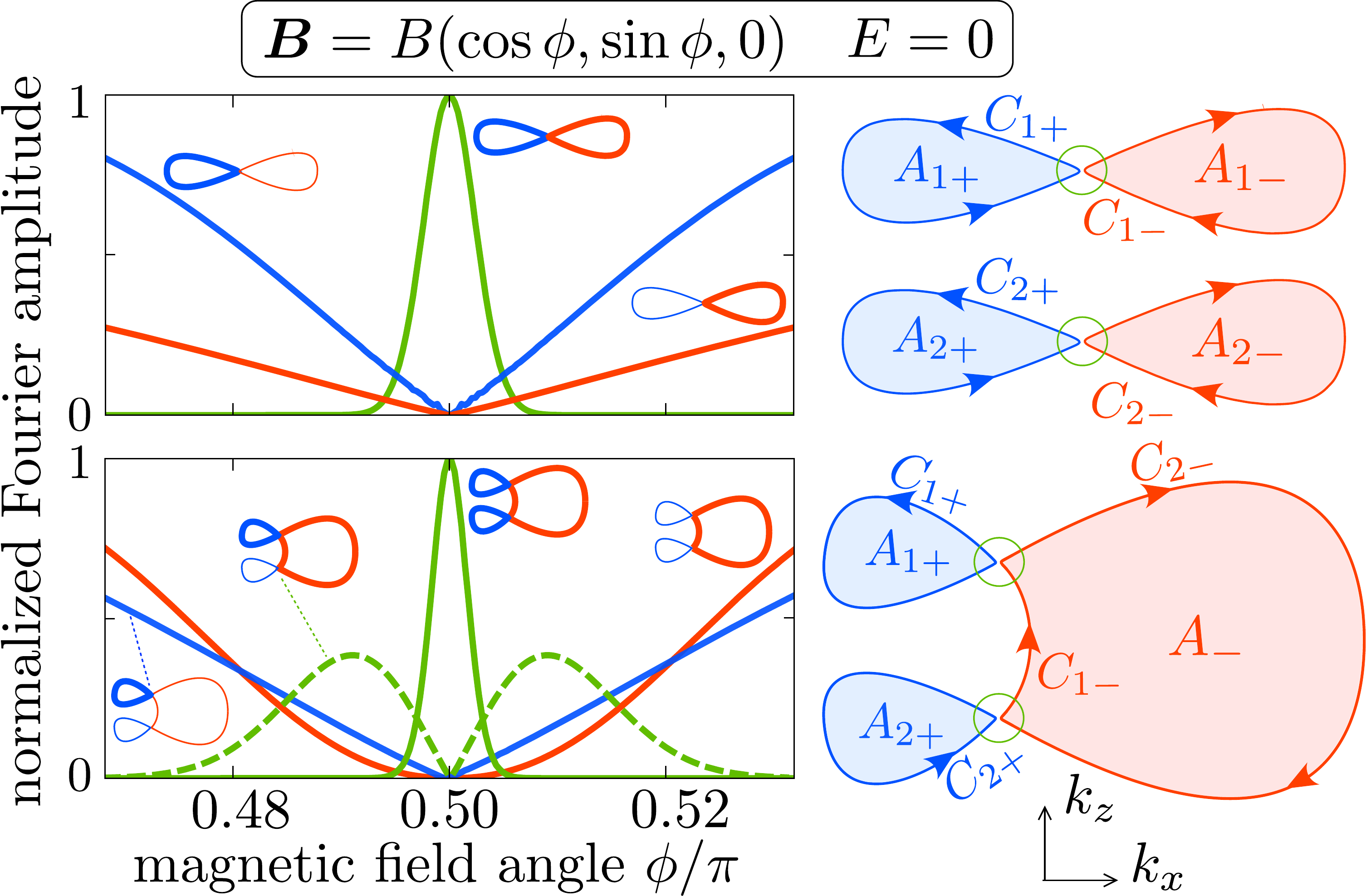}}
\caption{Left panels: Dependence on the orientation of the magnetic field of the amplitude of the magnetic quantum oscillations (normalized to unit maximal amplitude), for a fixed Fermi energy $E_{\rm F}=0$. Pairs of type-II Weyl points at $E=0$ with disconnected or connected Fermi surfaces are compared. The right panels show a cross-section through the electron and hole pockets. For each curve in the left panels the corresponding orbit is indicated. The calculations, detailed in App.\ C \cite{supplemental}, are for the Hamiltonian \eqref{Hmodel} with parameters $t=1$, $\mu=3$, $b=1.2$, $a_{\rm tilt}=1.7$ for all panels and $t'=2$, $\xi=0.08$ (top panels); $t'=1.7$,  $\xi=0.24$ (bottom panels).
\label{fig:exp_approx}
}
\end{figure}

If the loops $L_+=(\theta_+,\phi_+)$ and $L_-=(\theta_-,\phi_-)$ both wind around the M\"{o}bius strip, as they do in Fig.\ \ref{fig:magic_axis_fig}, they must necessarily intersect because of the twist. The point of intersection is the special axis at which both ${\cal C}_+$ and ${\cal C}_-$ touch the Weyl point. In App.\ B \cite{supplemental} we show that such non-contractible loops always exist if the Fermi surface is convex, independent of any symmetry requirement.

So far we considered Klein tunneling at a single Weyl point. A second Weyl point of opposite helicity necessarily exists in the Brillouin zone,  and this allows for topologically distinct Fermi surfaces \cite{Xu15}. In Fig.\ \ref{fig:exp_approx} we illustrate how Klein tunneling can distinguish connected from disconnected pairs of type-II Weyl cones, by the qualitatively different dependence on the magnetic field orientation.

Experimentally, Klein tunneling through a type-II Weyl point can be detected in measurements of the De Haas-Van Alphen effect in the magnetic susceptibility: If the magnetic axis is rotated towards the special alignment of Fig.\ \ref{fig:magic_axis_fig}, the high-frequency magnetic quantum oscillations from the electron and hole pockets would both be suppressed in favor of a low-frequency oscillation from the coupled orbits. The characteristic field for this magnetic breakdown would depend quadratically on the energy mismatch $E$ between the Weyl point and the Fermi energy, with unit tunnel probability in the limit $E\rightarrow 0$ as the defining signature of Klein tunneling in momentum space. With sufficient doping WTe$_2$ would produce disconnected type-II Weyl cones near the Fermi energy \cite{ref:Bernevig,note2}, while they are connected in undoped LaAlGe \cite{Xu16}. Klein tunneling is a powerful diagnostic for such topologically distinct Fermi surfaces.

We thank I. Adagideli, P. Baireuther, J. A. Hutasoit, B. Tarasinski, and M. Wimmer for valuable discussions. This research was supported by the Foundation for Fundamental Research on Matter (FOM), the Netherlands Organization for Scientific Research (NWO/OCW), and an ERC Synergy Grant.

\newpage

\appendix

\section{Low-energy limit of the four-band model Hamiltonian of a type-II Weyl semimetal}
\label{4band_derivation}

The dispersion along the $k_z$-axis (for $k_x=k_y=0$) of the four-band Hamiltonian \eqref{Hmodel} is given by
\begin{align}
E^{\rm high}_\pm &= \pm b \pm \sqrt{(2t-\mu)^2+t^2+2t(2t-\mu)\cos k_z}, \\
E^{\rm low}_\pm &= \pm b \mp \sqrt{(2t-\mu)^2+t^2+2t(2t-\mu)\cos k_z}.
\label{eq:Eofkz}
\end{align}
For $\mu>2t$ the two low-energy bands $E_\pm^{\rm low}$ form a pair of Weyl cones located at
\begin{align}
  K_z = \pm \arccos \left(\frac{(2t-\mu)^2+t^2-b^2}{2t(\mu-2t)}\right).
  \label{eq:Kz}
\end{align}
We wish to derive the corresponding low-energy Hamiltonian. Notice that for $k_x=k_y=0$ the Hamiltonian~\eqref{Hmodel} commutes with $\sigma_z$ and is thus block-diagonal. Each of the two blocks contains one low and one high energy band. At $\bm K = (0, 0, K_z)$ the corresponding low energy eigenstates are given by
\begin{align}
  \Psi^{\rm low}_+ = \frac{1}{N_+}\left(\frac{(2t-\mu)(2b-2t\sqrt{1-\cos^2K_z})}{(2t-\mu)^2-t^2+b^2}, 1, 0, 0\right)
  \label{eq:psip}, \\
  \Psi^{\rm low}_- = \frac{1}{N_-}\left(0, 0,-\frac{(2t-\mu)(2b+2t\sqrt{1-\cos^2K_z})}{(2t-\mu)^2-t^2+b^2}, 1\right).
  \label{eq:psim}
\end{align}
We expand the four-band Hamiltonian in the basis of these eigenstates:
\begin{widetext}
\begin{align}
  \left(\begin{array}{cc|cc}
	 \langle \Psi^{\rm low}_+ | H | \Psi^{\rm low}_+ \rangle & \langle \Psi^{\rm low}_+ | H | \Psi^{\rm low}_- \rangle & \langle \Psi^{\rm low}_+ | H | \Psi^{\rm high}_+ \rangle & \langle \Psi^{\rm low}_+ | H | \Psi^{\rm high}_- \rangle \\
	 \langle \Psi^{\rm low}_- | H | \Psi^{\rm low}_+ \rangle & \langle \Psi^{\rm low}_- | H | \Psi^{\rm low}_- \rangle & \langle \Psi^{\rm low}_- | H | \Psi^{\rm high}_+ \rangle & \langle \Psi^{\rm low}_- | H | \Psi^{\rm high}_- \rangle \\ \hline
	 \langle \Psi^{\rm high}_+ | H | \Psi^{\rm low}_+ \rangle & \langle \Psi^{\rm high}_+ | H | \Psi^{\rm low}_- \rangle & \langle \Psi^{\rm high}_+ | H | \Psi^{\rm high}_+ \rangle & \langle \Psi^{\rm high}_+ | H | \Psi^{\rm high}_- \rangle \\
	 \langle \Psi^{\rm high}_- | H | \Psi^{\rm low}_+ \rangle & \langle \Psi^{\rm high}_- | H | \Psi^{\rm low}_- \rangle & \langle \Psi^{\rm high}_- | H | \Psi^{\rm high}_+ \rangle & \langle \Psi^{\rm high}_- | H | \Psi^{\rm high}_- \rangle
  \end{array}\right) =
  \left( \begin{array}{c|c}
	 H_{\rm low} & V_{\rm high,low} \\ \hline
	 V_{\rm high,low}^\dagger & H_{\rm high}
  \end{array}\right).
  \label{eq:HKz}
\end{align}
\end{widetext}

At $\bm K$ the high and low energy blocks are uncoupled ($V_{\rm high,low}(0, 0, K_z) = 0$). Close to the Weyl point we have
\begin{align}
  \mathcal H\approx H_{\rm low} + V_{\rm high,low}^\dagger (H_{\rm high})^{-1}V_{\rm high,low}.
  \label{eq:SW}
\end{align}
Thus, to linear order in the deviation from $\bm K$ we can neglect this coupling and simply linearize $H_{\rm low}$. After some algebra we find the corresponding low-energy Weyl Hamiltonian,
\begin{align}
  \mathcal H = a_{\rm tilt} k_x \tilde\sigma_0 - v_x k_x \tilde\sigma_x + v_y k_y\tilde\sigma_y + v_z(k_z-K_z)\tilde\sigma_z.
  \label{eq:Hlow}
\end{align}
The matrices $\tilde\sigma_{0,x,y,z}$ are the identity and the Pauli matrices in the basis of $|\Psi^{\rm low}_\pm\rangle$. The anisotropic velocity components were given in the main text, Eq.\ \eqref{vxyzresult}.

\begin{figure}[tb]
\centerline{\includegraphics[width=0.8\linewidth]{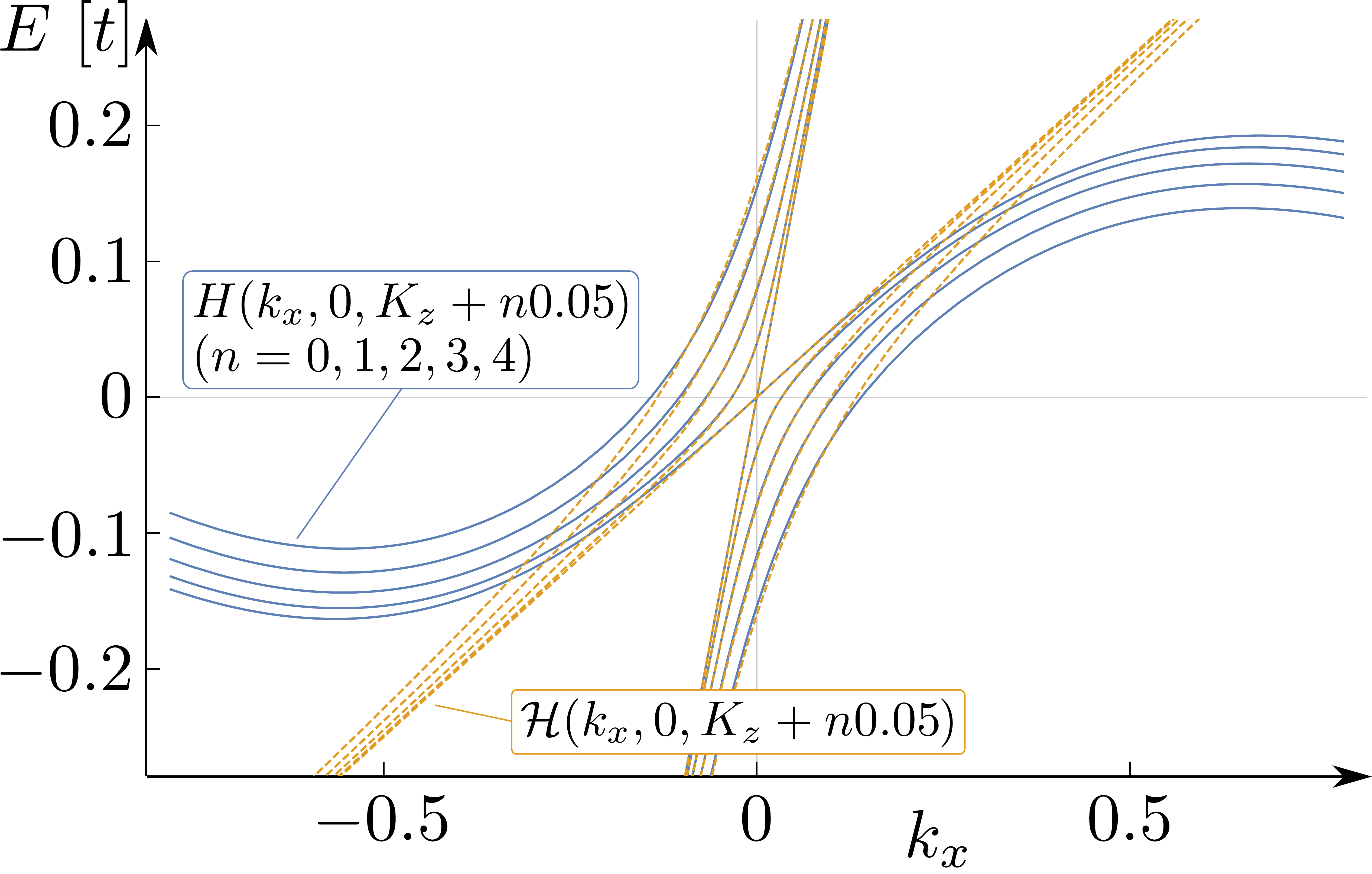}}
\caption{Dispersion close to a type-II Weyl point. Shown are five $k_z$-subbands at $k_y=0$; blue-solid lines show the dispersion of the four-band model~\eqref{Hmodel}; yellow-dashed lines show the corresponding low energy description~\eqref{eq:Hlow}. Parameters are the same as in  Fig.\ \ref{fig:Landau_levels}.}
\label{fig:low_energy}
\end{figure}

Fig.\ \ref{fig:low_energy} shows a comparison between the type-II Weyl cone of the four-band model and its effective low-energy description.

\section{Topological protection of the special magnetic field axis for Klein tunneling between electron and hole pockets}
\label{app:magic_axis_proof}

\begin{figure}[tb]
\centerline{\includegraphics[width=0.8\linewidth]{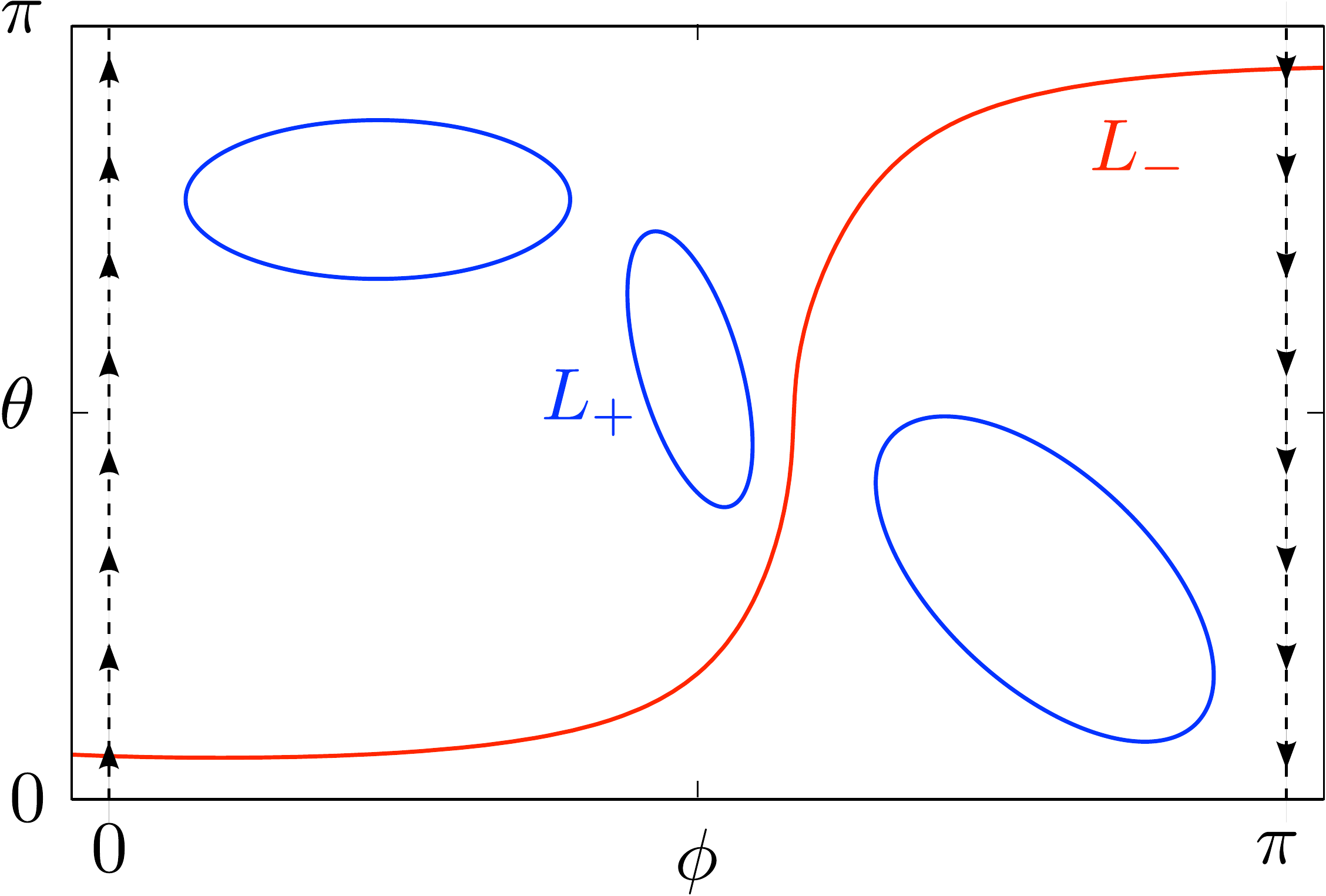}}
\caption{Same as Fig.\ \ref{fig:magic_axis_fig}, but now the incontractible loop $L_+$ is replaced by a set of contractible loops, containing the entire set of magnetic field axes with extremal contours in the electron pocket that touch the Weyl point. This arrangement would avoid the topological protection of the intersection of incontractible loops in a M\"{o}bius strip, but we show by contradiction that it cannot happen in a convex electron pocket.
\label{fig:magic_axis_fig_app}
}
\end{figure}

The topology of the M\"{o}bius strip (the projective plane ${\mathbb P}_2$) protects the intersection of two incontractible loops, ensuring the existence of a special magnetic field axis where the extremal contours ${\cal C}_\pm$ in the electron and hole pockets both touch the Weyl point at $E=0$. This is the arrangement shown in Fig.\ \ref{fig:magic_axis_fig}. Contractible loops can avoid the intersection, as they do in Fig.\ \ref{fig:magic_axis_fig_app}. For convex electron and hole Fermi surfaces the existence of incontractable loops is guaranteed by the following argument.

Consider the full set ${\cal S}_+$ of magnetic field axes for which the extremal contour ${\cal C}_+$ in the electron pocket touches the Weyl point. If this set would consist only of contractible loops, then we would be able to pass an incontractible loop $L$ through $\mathbb{P}_2$ that avoids ${\cal S}_+$. We will now see that this leads to a contradiction.

For a convex Fermi surface each field axis $\hat{\bm{n}}$ on $L$ is associated with a unique extremal contour $C(\hat{\bm{n}})$. By construction, the contour $C(\hat{\bm{n}})$ lies in a plane normal to $\hat{\bm{n}}$. The direction $\hat{\bm{n}}$ defines whether the Weyl point lies above or below this plane. Inversion of the axis produces the same extremal contour and therefore the same normal plane, with ``above'' and ``below'' interchanged. As we follow the incontractible loop $L$ from polar angle $\phi=0$ to $\phi=\pi$, the field axis is inverted, so at some axis $\hat{\bm{n}}_0$ on $L$ the Weyl point must move from above to below the plane. As motion of the plane is continuous, this can only happen if the Weyl point actually lies on $C(\hat{\bm{n}}_0)\in L$. This would mean that $C(\hat{\bm{n}}_0)\in {\cal S}_+$, which we had excluded by the construction of $L$.

The same argument can be applied to the hole pocket, and we conclude that for both the (convex) electron and hole pockets there must exist incontractible loops $L_\pm$ of field axes with extremal contours that touch the Weyl point.

\section{Klein tunneling for pairs of connected type-II Weyl points}
\label{app_angular_dep}

The curves in Fig.\ \ref{fig:exp_approx} are calculated as follows. The Fermi level is fixed at the energy $E=0$ of the Weyl points and the magnetic field $\bm{B}$ is rotated in the $x$--$y$ plane, staying close to the $y$-axis (angles $\theta=0$, $|\phi/\pi-1/2|\ll 1$). We assume that the dominant $\phi$-dependence of the amplitude of the magnetic quantum oscillations is then given by the Klein tunneling probability. 

For a given field orientation $\hat{\bm{n}}$ we define $T(q)$ as the Klein tunneling probability between electron and hole pockets at $E=0$ and $\hat{\bm{n}}\cdot\bm{k}=q$. Because of the symmetry of our band structure, both Weyl points have the same $T$. We then take a planar cross-section of the Fermi surface at a momentum $q$ parallel to the field and select one of the contours indicated in the left panels of Fig.\ \ref{fig:exp_approx}. The contour encloses a signed area $A(q)$ and we determine the $q_c$ at which the area is extremal, $A'(q_c)=0$. We calculate $T_c=T(q_c)$ using the general Landau-Zener formula \eqref{TLZ_general}. Finally, we follow the contour for one period, collecting a factor $\sqrt{T_c}$ for each transmission through a Weyl point and a factor $\sqrt{1-T_c}$ for each reflection. The product of these factors is plotted in Fig.\ \ref{fig:exp_approx} (left panels) as a function of the field orientation $\phi$.


\begin{thebibliography}{99}
\bibitem{Wey29} H. Weyl, Z. Phys. \textbf{56}, 330 (1929).
\bibitem{Her37} C. Herring, Phys. Rev. \textbf{52}, 365 (1937).
\bibitem{Volovik} G. E. Volovik, \textit{The Universe in a Helium Droplet} (Oxford University Press, 2003).
\bibitem{Huh02} P. Huhtala and G. E. Volovik, JETP \textbf{94}, 853 (2002).
\bibitem{Kaw12} T. Kawarabayashi, Y. Hatsugai, T. Morimoto, and H. Aoki, Int. J. Mod. Phys. Conf. Series \textbf{11}, 145 (2012).
\bibitem{Tre15} M. Trescher, B. Sbierski, P. W. Brouwer, and E. J. Bergholtz, Phys. Rev. B \textbf{91}, 115135 (2015).
\bibitem{Sun15} Y. Sun, S.-C. Wu, M. N. Ali, C. Felser, and B. Yan, Phys. Rev. B \textbf{92}, 161107(R) (2015).
\bibitem{Xu15} Y. Xu, F. Zhang, and C. Zhang, Phys. Rev. Lett. \textbf{115}, 265304 (2015).
\bibitem{ref:Bernevig} A. A. Soluyanov, D. Gresch, Z. Wang, Q-S. Wu, M. Troyer, X. Dai, and B. A. Bernevig, Nature {\bf 527}, 495 (2015).
\bibitem{Aut16} G. Aut\'{e}s, D. Gresch, A. A. Soluyanov, M. Troyer, and O. V. Yazyev, arXiv:1603.04624.
\bibitem{Koe16} K. Koepernik, D. Kasinathan, D. V. Efremov, S. Khim, S. Borisenko, B. B\"{u}chner, and J. van den Brink, arXiv:1603.04323.
\bibitem{Mue16} L. Muechler, A. Alexandradinata, T. Neupert, and R. Car, arXiv:1604.01398.
\bibitem{Goe08} M. O. Goerbig, J.-N. Fuchs, G. Montambaux, F. Piechon, Phys. Rev. B \textbf{78}, 045415 (2008).
\bibitem{Hua16} L. Huang, T. M. McCormick, M. Ochi, Z. Zhao, M. Suzuki, R. Arita, Y. Wu, D. Mou, H. Cao, J. Yan, N. Trivedi, and A. Kaminski, arXiv:1603.06482.
\bibitem{Xu16} S.-Y. Xu, N. Alidoust, G. Chang, H. Lu, B. Singh, I. Belopolski, D. Sanchez, X. Zhang, G. Bian, H. Zheng, M.-A. Husanu, Y. Bian, S.-M. Huang, C.-H. Hsu, T.-R. Chang, H.-T. Jeng, A. Bansil, V. N. Strocov, H. Lin, S. Jia, and M. Z. Hasan, arXiv:1603.07318.
\bibitem{Den16} K. Deng, G. Wan, P. Deng, K. Zhang, S. Ding, E. Wang, M. Yan, H. Huang, H. Zhang, Z. Xu, J. Denlinger, A. Fedorov, H. Yang, W. Duan, H. Yao, Y. Wu, S. Fan, H. Zhang, X. Chen, and S. Zhou, arXiv:1603.08508.
\bibitem{Jia16} J. Jiang, Z. K. Liu, Y. Sun, H. F. Yang, R. Rajamathi, Y. P. Qi, L. X. Yang, C. Chen, H. Peng, C.-C. Hwang, S. Z. Sun, S.-K. Mo, I. Vobornik, J. Fujii, S. S. P. Parkin, C. Felser, B. H. Yan, and Y. L. Chen, arXiv:1604.00139.
\bibitem{Lia16} A. Liang, J. Huang, S. Nie, Y. Ding, Q. Gao, C. Hu, S. He, Y. Zhang, C. Wang, B. Shen, J. Liu, P. Ai, L. Yu, X. Sun, W. Zhao, S. Lv, D. Liu, C. Li, Y. Zhang, Y. Hu, Y. Xu, L. Zhao, G. Liu, Z. Mao, X. Jia, F. Zhang, S. Zhang, F. Yang, Z. Wang, Q. Peng, H. Weng, X. Dai, Z. Fang, Z. Xu, C. Chen, and X. J. Zhou, arXiv:1604.01706.
\bibitem{NXu16} N. Xu, Z. J. Wang, A. P. Weber, A. Magrez, P. Bugnon, H. Berger, C. E. Matt, J. Z. Ma, B. B. Fu, B. Q. Lv, N. C. Plumb, M. Radovic, E. Pomjakushina, K. Conder, T. Qian, J. H. Dil, J. Mesot, H. Ding, and M. Shi, arXiv:1604.02116.
\bibitem{Wang16} C. Wang, Y. Zhang, J. Huang, S. Nie, G. Liu, A. Liang, Y. Zhang, B. Shen, J. Liu, C. Hu, Y. Ding, D. Liu, Y. Hu, S. He, L. Zhao, L. Yu, J. Hu, J. Wei, Z. Mao, Y. Shi, X. Jia, F. Zhang, S. Zhang, F. Yang, Z. Wang, Q. Peng, H. Weng, X. Dai, Z. Fang, Z. Xu, C. Chen, and X. J. Zhou, arXiv:1604.04218.
\bibitem{Har09} N. Harrison and S. E. Sebastian, Phys. Rev. B \textbf{80}, 224512 (2009).
\bibitem{Cha15} M.-C. Chang and M.-F. Yang, Phys. Rev. B \textbf{92}, 205201 (2015).
\bibitem{Zyu16} A. A. Zyuzin and R. P. Tiwari, arXiv:1601.00890.
\bibitem{Yu16} Z. Yu, Y. Yao, and S. A. Yang, arXiv:1604.04030.
\bibitem{ref:Ashcroft_Mermin} N. W. Ashcroft and N. D. Mermin, \emph{Solid State Physics} (Cengage Learning, 2011).
\bibitem{Wan16} C. M. Wang, H.-Z. Lu, and S.-Q. Shen, arXiv:1604.01681.
\bibitem{note1} For a convex electron or hole pocket one has ${\rm sign}\,A''_\pm(q_c)=\mp 1$ so the factor $[-iA''_\pm(q_c)]^{-1/2}$ in Eq.\ \eqref{deltarho} contributes a phase shift $\mp\pi/4$ to the magnetic oscillations.
\bibitem{ref:Onsager} L. Onsager, Phil. Mag. {\bf 43}, 1006 (1952).
\bibitem{ref:Shoenberg} D. Shoenberg, \emph{Magnetic Oscillations in Metals} (Cambridge University Press, 1984).
\bibitem{Bee08} C. W. J. Beenakker, Rev. Mod. Phys. \textbf{80}, 1337 (2008).
\bibitem{Nan11} R. Nandkishore and L. Levitov, PNAS \textbf{108}, 14021 (2011).
\bibitem{Lan77} L. D. Landau and E. M. Lifshitz, \textit{Course of Theoretical Physics}, Vol.\ 3 (Elsevier, Oxford, 1977).
\bibitem{ref:Blount} E. I. Blount, Phys. Rev. {\bf 126}, 1636 (1962).
\bibitem{ref:Winkler} S. Keppeler and R. Winkler, Phys. Rev. Let. {\bf 88}, 046401 (2002).
\bibitem{Vazifeh2013} M. M. Vazifeh and M. Franz, Phys. Rev. Lett. {\bf 111}, 027201 (2013). The Hamiltonian \eqref{Hmodel} differs from that in this reference by the addition of the $a_{\rm tilt}$ and $\xi$ terms, and also in the replacement of $\tau_y\sigma_0\sin k_z$ by $\tau_z\sigma_0\sin k_z$. This last change was introduced to produce small electron and hole pockets that do not spread out over the entire Brillouin zone.
\bibitem{supplemental} For the appendices see the supplemental material.
\bibitem{kwant} To discretize the model Hamiltonian \eqref{Hmodel} we used the {\sc kwant} toolbox: C. W. Groth, M. Wimmer, A. R. Akhmerov, and X. Waintal, New J. Phys. \textbf{16}, 063065 (2014).
\bibitem{note2} Z. Zhu, X. Lin, J. Liu, B. Fauqu\'{e}, Q. Tao, C. Yang, Y. Shi, and K. Behnia, Phys. Rev. Lett. \textbf{114}, 176601 (2015). This recent experimental study of magnetic quantum oscillations in semimetallic WTe$_2$ reports the sudden appearance of a new frequency above a critical field. Since the new frequency is the sum rather than the difference of the low-field frequencies, it cannot be associated with Klein tunneling through a Weyl point. 
\end{thebibliography}
\end{document}